\documentclass[preprint]{aastex62}
\received{}
\revised{}
\accepted{}
\submitjournal{ApJ}

\shorttitle{Untwisting of a solar filament}
\shortauthors{Chen et al.}

\begin{document}

\title{Untwisting and Disintegration of a Solar Filament Associated with Photospheric Flux Cancellation}

\correspondingauthor{Huadong Chen}
\email{hdchen@nao.cas.cn}

\author{Huadong Chen}
\affil{CAS Key Laboratory of Solar Activity, 
     National Astronomical Observatories, 
     Chinese Academy of Sciences, 
      Beijing 100101, People's Republic of China}
\affil{School of Astronomy and Space Science, University of Chinese Academy of Sciences, Beijing 100049, People's Republic of China}

\author{Ruisheng Zheng}
\affil{Shandong Provincial Key Laboratory of Optical Astronomy 
and Solar-Terrestrial Environment, and Institute of Space Sciences, 
Shandong University, Weihai 264209, People's Republic of China}

\author{Leping Li}
\affil{CAS Key Laboratory of Solar Activity, 
     National Astronomical Observatories, 
     Chinese Academy of Sciences, 
      Beijing 100101, People's Republic of China}
      \affil{School of Astronomy and Space Science, University of Chinese Academy of Sciences, Beijing 100049, People's Republic of China}

\author{Suli Ma}
\affiliation{College of Science, China University 
of Petroleum, Qingdao 266580, People's Republic of China}
\affiliation{CAS Key Laboratory of Solar Activity, 
     National Astronomical Observatories, 
     Chinese Academy of Sciences, 
      Beijing 100101, People's Republic of China}

\author{Yi Bi}
\affil{Yunnan Astronomical Observatory, 
Chinese Academy of Sciences, 
Kunming 650011, People's Republic of China}
\affil{CAS Key Laboratory of Solar Activity, 
     National Astronomical Observatories, 
     Chinese Academy of Sciences, 
      Beijing 100101, People's Republic of China}

\author{Shuhong Yang}
\affil{CAS Key Laboratory of Solar Activity, 
     National Astronomical Observatories, 
     Chinese Academy of Sciences, 
      Beijing 100101, People's Republic of China}
      \affil{School of Astronomy and Space Science, University of Chinese Academy of Sciences, Beijing 100049, People's Republic of China}



\begin{abstract}
Using the high-resolution observations from New Vacuum Solar Telescope (NVST) jointly with the $Solar~Dynamics~Observatory$ data, we investigate two successive confined eruptions (Erup1 and Erup2) of a small filament in a decaying active region on 2017 November 10. 
During the process of Erup1, the overlying magnetic arcade is observed to inflate with the rising filament at beginning and then stop the ongoing of the explosion.
In the hot EUV channel, a coronal sigmoidal structure appears during the first eruption and fade away after the second one.
The untwisting rotation and disintegration of the filament in Erup2 are clearly revealed by the NVST H$\alpha$ intensity data, hinting at a pre-existing twisted configuration of the filament.
By tracking two rotating features in the filament, the average rotational angular velocity of the unwinding filament is found to be $\sim$10.5\degr\ min$^{-1}$.
A total twist of $\sim$1.3 $\pi$ is estimated to be stored in the filament before the eruption, which is far below the criteria for kink instability.
In the course of several hours prior to the event, some photospheric flux activities, including the flux convergence and cancellation, are detected around the northern end of the filament, where some small-scale EUV brightenings are also captured. 
Moreover, strongly-sheared transverse fields are found in the cancelling magnetic features from the vector magnetograms.
Our observational results support the flux cancellation model, in which the interaction between the converging and sheared opposite-polarity fluxes destabilizes the filament and triggers the ensuing ejection.
\end{abstract}



\section{Introduction} \label{sec:intro}
Solar filaments, also known as prominences, consist of relatively cool, dense plasma that is suspended in the hot tenuous solar corona.
They are always located above polarity inversion lines (PILs) separating opposite polarities of radial fields in the photosphere (Babcock \& Babcock 1955), which can be inside or at the border of active regions (ARs) or on the quiet Sun.
Substantial observed characteristics, including the direct magnetic field measurements of filaments (Leroy 1989; Paletou \& Aulanier 2003) reflect that they are present in highly non-potential magnetic structures (e.g., Tandberg-Hanssen 1995; D\'{e}moulin 1998; Labrosse et al. 2010; Mackay et al. 2010).
However, the basic magnetic configuration of the filament and its relationship with the surrounding coronal structures are still under debate (e.g., Berger et al. 2011; Yang et al. 2014a; Cheng et al. 2014b; Shen et al. 2015).
It has been commonly accepted that the heavy filament material is supported by magnetic tension force from the dipped coronal fields (Kiepenheuer 1953), which could have normal or inverse polarity (Kippenhahn \& Schl\"{u}ter 1957; Kuperus \& Raadu 1974; Leroy et al. 1984; Bommier \& Leroy 1998), i.e., the component of magnetic field perpendicular to the body of filament has a direction same or opposite to that of the potential field.

In the flux rope theoretical model (e.g., Malherbe \& Priest 1983; D\'{e}moulin \& Priest 1989; Aulanier \& D\'{e}moulin 1998; Amari et al. 1999; van Ballegooijen et al. 2000), the inverse polarity configuration means that filament is embedded in a helical or twisted field structure, which stores much more energy than the normal polarity configuration.
Even in the three-dimensional sheared arcade model, as proposed by DeVore \& Antiochos (2000) and Aulanier et al. (2002), such a twisted field geometry of filaments is also expected to form due to the reconnection between the sheared and external field at a large shear condition.
On the assumption of frozen-in condition (due to the high electrical conductivity of solar atmosphere), the fine thread structures of filament matter probably reflect the magnetic topology therein.
Helical twisted thread structures have been directly observed in both quiescent and eruptive prominences (e.g., Vr\v{s}nak et al. 1988, 1991; Dere et al. 1999).
Sometimes, filaments or jets exhibit an apparent unwinding rotational motions during their ejections (e.g., Vr\v{s}nak et al. 1991; Shen et al. 2011b; Hong et al. 2013; Chen et al. 2012, 2017; Yan et al. 2014a, 2014b; Yang et al. 2015), also suggesting a very likely pre-existing twisted magnetic topology of the ejected structures.
In the case studied by Bi et al. (2013), they proposed that the rotation of the eruptive filament might originate from the action of the asymmetric deflection, which was caused by the interaction between the erupting and the surrounding magnetic fields.


It has been well established that a flux tube will be subject to ideal magnetohydrodynamic (MHD) kink instability when the magnetic twist in the flux tube exceeds a critical value $\Phi_{c}$ (e.g., Raadu 1972; Hood \& Priest 1979; Einaudi \& Van Hoven 1983; Velli et al. 1990; Baty 2001).
The kink instability of a coronal magnetic flux rope can trigger not only confined but also ejective solar eruptions.
The decrease of the overlying field with height plays an important role in deciding whether the event is a confined or an ejective one (T\"{o}r\"{o}k \& Kliem 2005), which involves another MHD instability -- torus instability (e.g., Bateman 1978; Kliem \& T\"{o}r\"{o}k 2006).
For a straight, cylindrically symmetric flux tube, its total twist can be expressed as, 
\begin{equation}
\Phi=\frac{lB_{\phi}(r)}{rB_{z}(r)},
\end{equation}
where $l$ is the length, $r$ is the minor radius, and $B_{\phi}$ and $B_{z}$ are the azimuthal and axial field components, respectively.
The critical value for determining stability or instability, $\Phi_{c}$, rests on the details of the considered flux system in the model, such as the aspect ratio of the loop, the effect of line tying, the ratio of the plasma to magnetic pressure, the radial profile of the twist, and the stabilizing influence of the overlying magnetic arcade.
Investigations of numerical calculations and MHD simulations have revealed that a typical value of $\Phi_{c}$ ranges from 2.5$\pi$ to 6$\pi$ for a straight cylindrical or an arched, line-tied twisted flux tube (Hood \& Priest 1979, 1981; Miki\'{c} et al. 1990; Van Hoven et al. 1995; T\"{o}r\"{o}k \& Kliem 2003; T\"{o}r\"{o}k et al. 2004, 2014).

Besides ideal MHD instabilities, there are also many other mechanisms suggested to be triggers of solar eruptions, such as sunspot rotation (e.g., Amari et al. 1996; Zhang et al. 2007; Yan \& Qu 2007; Yan et al. 2009), twisting overlying field (T\"{o}r\"{o}k et al. 2013); shearing of magnetic arcade (e.g., Miki\'{c} et al. 1988), tether cutting (e.g., Moore \& Roumeliotis 1992; Moore et al. 2001; Liu et al. 2010; Chen et al. 2014), magnetic breakout (e.g., Antiochos et al. 1999; Chen et al. 2016b), flux feeding or injection (e.g., Chen 1989, 1996; Liu et al. 2012; Zhang et al. 2014; Kliem et al. 2014), and flux emergence or cancellation (e.g., Heyvaerts et al. 1977; Livi et al. 1989; van Ballegooijen \& Martens 1989; Wang \& Shi 1993; Feynman \& Martin 1995; Wang \& Sheeley 1999; Chen \& Shibata 2000; Lin et al. 2001; Jing et al. 2004; Inhester et al. 1992; Amari et al. 2003a; Roussev et al. 2004; Zuccarello et al. 2012) etc, as recently reviewed by Chen (2011) and Green et al. (2018).
According to the definition by Livi et al. (1985) and Martin et al. (1985), flux cancellation refers to the mutual disappearance of photospheric magnetic flux in closely spaced features of opposite polarity, which can be observed throughout the quiet sun, at the edge of or in ARs and sometimes may take place concurrently with the flux emergence (Feynman \& Martin 1995; Wang \& Sheeley 1999; Chen et al. 2008; Louis et al. 2014; Li et al. 2015).
The likely theoretical interpretations on the flux disappearance during cancellation mainly embrace submergence of small $\Omega$-like loop (e.g., Harvey et al. 1999), upward expulsion of U-shaped flux tube (e.g., Spruit et al. 1987), and annihilation of magnetic flux (e.g., Martin et al. 1985; Amari et al. 2010).
A prevailing thought is that these processes result from the reconnection of the cancelling magnetic components (e.g., Zwaan 1978, 1987; Wang \& Shi 1993).

The reconnection associated with flux emergence or cancellation may play two important roles in prompting solar eruptions. On one hand, it triggers an explosion by modifying the configuration of the overlying field restraining the filament or flux rope, as modeled and simulated by Chen \& Shibata (2000), Lin et al. (2001), and Kusano et al. (2012) and supported by the observations of Wang \& Sheeley (1999), Jiang et al. (2007), Li et al. (2015), and Louis et al. (2015).  
In this case, it could also be the tether-cutting reconnection occurring between the strongly sheared core fields, which was first suggested by Moore \& Labonte (1980) and Moore et al. (2001) and then reported by many other works (e.g., Sterling \& Moore 2005; Yurchyshyn et al. 2006; Sterling et al. 2007; Kim et al. 2008; Green \& Kliem 2009; Liu et al. 2010; Chen et al. 2014, 2015, 2016a; Yang et al. 2016).
On the other, flux cancellation is a likely formation mechanism of sigmoidal flux ropes (van Ballegooijen \& Martens 1989; Green et al. 2011; Savcheva et al. 2012; Yardley et al. 2016). 
It can convert the sheared arcade field into the helical field of the rope, increase the magnetic pressure and eventually lead to the loss of equilibrium of the system at some critical point coupled with the effect of MHD instabilties (Forbes et al. 2006; Su et al. 2011).
This scenario has been investigated by numerical simulations (e.g., Amari et al. 2003a, 2003b; Roussev et al. 2004; Fang et al. 2012; Zuccarello et al. 2012) and supported by some observational works (e.g., Green et al. 2011; Savcheva et al. 2012; Chen et al. 2014; Yan et al. 2015; Yardley et al. 2016).
It is worth pointing out that the two roles of flux cancellation in motivating eruptions should be not mutually exclusive and they may work cooperatively in a single event (e.g., Sterling et al. 2011; Chen et al. 2014).

Up to now, the initiation mechanism of solar eruptions is still not sufficiently understood.
In addition to numerical study, more high-resolution observations are needed to present the real magnetic configuration of the erupting system.
This would be useful to elucidate the precise origin of the eruption, which is of key importance to the forecasting of space weather.
In this work, we use the high-resolution observations from New Vacuum Solar Telescope (NVST; Liu et al. 2014) together with the $Solar~Dynamics~Observatory$ ($SDO$; Pesnell et al. 2012) data to scrutinize two successive confined eruptions of a small filament occurring in a decaying AR.
The untwisting rotation and disintegration of the erupting filament are disclosed by the NVST H$\alpha$ line-center and off-band intensity data, reflecting a preceding twisted structure of the filament before its eruption.
However, the results from a detailed calculation indicate that the twisted filament was not subjected to the kink instability probably.
According to the evolution of the associated photospheric magnetic flux, we suggest that the flux convergence and cancellation detected in the vicinity of one end of the filament may play a main role in triggering the eruptions.
The remaining part of this article is organized as follows: In Section~2 we describe the observational data used in our study. In Section~3 we present the detailed evolutions of the successive filament eruptions (Section~3.1 and 3.2) and the associated photospheric magnetic flux (Section~3.3). We finally summary and discuss our results in Section~4.

\section{Observations}
On 2017 November 10, a solar filament with a length of $\sim$17 Mm successively erupted for two times during $\sim$1 hour in a small decaying AR near the solar disk's center ($\sim$S10E05).
According to the observation from the Large Angle and Spectrometric Coronagraph (LASCO; Brueckner et al. 1995) on $Solar~and~Heliospheric~Observatory$ ($SOHO$), no coronal mass ejection relates to this event.
The two confined eruptions mainly took place during the periods of 02:07--02:25 UT (Erup1) and 02:40--03:10 UT (Erup2), respectively.
The Atmospheric Imaging Assembly (AIA; Lemen et al. 2012) on board $SDO$ covered this event well and provides us full-disk images up to 0.5 $R_\sun$ above the solar limb with 0\farcs6  pixel size and 12 s cadence in 10 wavelengths.
We mainly used the data (Level 1.5 images) at 5 EUV channels centered at 
304 \AA\ (\ion{He}{2}, 0.05 MK),
171 \AA\ (\ion{Fe}{9}, 0.6 MK), 
211 \AA\ (\ion{Fe}{14}, 2 MK), 
335\AA\ (\ion{Fe}{16}, 2.5 MK), 
and 94 \AA\ (\ion{Fe}{18}, 7 MK), 
respectively.

The observations of NVST started from 02:24 UT on that day and only captured 
Erup2. 
The H$\alpha$ line-center and off-band (6562.8$\pm$0.3 \AA) intensity data
from NVST have a pixel size of 0\farcs163  and a cadence of $\sim$36 s.
We used the Level 1+ data, which had been calibrated through dark current subtraction 
and flat field correction, and were generated through reconstructing the data
by speckle masking (Weigelt 1977; Lohmann et al. 1983).
By comparing some identical features in the AIA EUV and NVST H$\alpha$ intensity 
images, such as bright points, fibrils, or chromospheric plages, we have co-aligned 
the AIA and NVST data.
The accuracies of all co-alignments are better than 1\arcsec .
We also utilized the full-disk longitudinal and vector magnetograms with a 0\farcs5 plate scale from the Helioseismic and Magnetic Imager (HMI; Schou et al. 2012) on board $SDO$ to analyze the evolution of photospheric magnetic fluxes near the filament footpoints.
They were produced at a cadence of 45 s and 12 minute, respectively (Hoeksema et al. 2014).
The HMI vector magnetic field data were already disambiguated using 
the standard procedures in the Solar SoftWare (SSW, Freeland \& Handy 1998).

\section{Results}
\subsection{Successive Filament Eruptions in the AIA EUV Channel}
The host AR started to emerged from November 7 and peaked at the end of November 8. When the event occurred on November 10, the AR had been in its decaying phase.
The AIA 171 \AA\ images in Figure~1 present the main processes of the two confined eruptions.
Before Erup1, the filament was basically aligned along the north-south direction in the AR. 
At about 02:07 UT, an EUV brightening appeared near the northern end of the filament and thereafter the eruption commenced.
The filament first rose up to a (projected) height of $\sim$5 Mm, and then partially fell back to the surface, while a different part of the filament was expelled toward the south (roughly following the magnetic neutral line) (see panels (a3) and (a4)). 
Meanwhile, the overlying arcade structure (labeled ``Ar'' in Figure~1(a2)) of the AR remained intact throughout the event; that is, the eruption did not burst through that overlying field.
After Erup1, the filament appeared again at its former location (see Figure~1 (b1)).

At 02:40 UT, 15 minutes after the end of Erup1, similar EUV brightenings
arose near the northern footpoint of the filament.
Then, the filament broke away from this end and bifurcated into several threads,
as its southern end remained anchored onto the photosphere.
Due to the small scale of the filament as well as the overlying coronal emission from 
the hot EUV line, the detailed dynamical evolution of the filament during Erup2 was not clearly
revealed by the AIA data.
But we can still detect the rapid rotation of the southern root of the filament 
during the eruption (see Figure~1(b4) and the movie~1).
After 03:05 UT, the filament gradually disappeared in the EUV lines.

Our observations reflect that the overlying background field of the AR probably played an important role in confining the eruptions.
In this situation, the strong downward magnetic tension from Ar stopped the erupting filament material escaping from the solar atmosphere.
According to the theory of MHD instability, we predict that the overlying restraining field's  strength did not fall off fast enough with the height so as to allow the occurrence of torus instability (Kliem \& T\"{o}r\"{o}k 2006; Aulanier et al. 2010).
Similar processes have been reported by some observational works (e.g., Ji et al. 2003, Guo et al. 2010; Cheng et al. 2011; Shen et al. 2011a; Zheng et al. 2012; Chen et al. 2013; Song et al. 2014; Yang et al. 2014b; Li et al. 2018; Ning et al. 2018) and numerically simulated by T\"{o}r\"{o}k \& Kliem (2005) and Fan \& Gibson (2007).


An interesting phenomenon is found in the hot AIA 211 \AA\ channel that a transient
sigmoid structure surrounding the filament appeared during Erup1 (Figure~2).
Generally, coronal sigmoids are the observational signatures of sheared and/or twisted fields (e.g., Rust \& Kumar 1996; Green \& Kliem 2009; Green et al. 2011; Savcheva et al. 2012).
Two possibilities are responsible for the appearance of the sigmoid.
Firstly, the sigmoid had existed before eruption and just brightened due to the reconnection heating during the eruption.
Secondly, the S-shape formed in the course of eruption might have been due to the reconnection of sheared fields (Moore et al. 2001) or the injection of the poloidal field (Priest 2014).
Since there were also some activities taking place in the AR several hours before the event but no similar S-shaped configuration arose, we incline to the second situation.
The sigmoid faded away after Erup2 and did not come into being again, implying a permanent destruction or change of its structure during Erup2.
In the hotter AIA 335 \AA\ and 94 \AA\ passbands, the observational signals of the bright sigmoid were detected to be faint.
These results indicate that the plasma in the sigmoid primarily had a temperature as high as $\sim$2 MK, which is the central formation temperature of the AIA 211 \AA\ emission.




\subsection{Erup2 Observed by NVST} \label{subsec:}
In the studied event, only the second confined eruption Erup2 was captured by NVST.
The NVST H$\alpha$ line-center intensity data (Figure~3(a1)--(a6)) of higher spatial resolution clearly revealed that the filament underwent an untwisting motion during Erup2 (also see the movie~2). 
From Figure~3(a1) to (a2), we can see that some filament threads formerly twined 
around together split into several strands due to unwinding.
Subsequently, the untwisting motion developed southward along the filament
and the southern end began to rotate anticlockwise.
As a result, the filament threads kept loosening from tight state.
After $\sim$02:48 UT, it can be observed that some threads stripped away from the main structure and they proceeded to spin and unwind from each other until $\sim$03:03 UT, as indicated by the thick and curved arrows in Figure~3(a3)--(a5).

The panels (b1)--(b3) of Figure~3 are H$\alpha$ Doppler intensity images,
which were obtained by subtracting the H$\alpha$ blue-wing data from the red-wing data.
So, as for the filament, the dark (bright) features in the Doppler images correspond
to the parts with redshifts (blueshifts).
From Figure~3(b1), it can be seen that the filament spine exhibited a twisted structure with blueshift while the two legs appeared as redshifted features, implying that some twist had been stored in the filament and the filament experienced a slow rise prior to the eruption.
The redshift characteristics of the filament legs may be related to the downward flows along the field lines.
When the filament was unwinding after the commencement of Erup2, we can see
that the uniform red- or blueshift signature arose in the same thread structure (Figure~3(b2)).
It is consistent with the scenario that the plasma should have similar dynamics in one strand of the filament if considering the freezing-in effect of the plasma-magnetic field coupling.
In Figure~3(b3), we can see that as the southern end of the filament rotated, the leg connecting with it formed a helical structure and showed ejective characteristic.

The untwisting motion of the filament presented here reflects that some twist may have been kept in the filament before its eruption.
An important question arises regarding how much the total stored twist was, the answer to which will provide a clue for clarifying the triggering mechanism of the filament eruptions in this event.
To get the answer, the rotational motion of the filament was first investigated in detail.
We chose two apparent rotating features (``RF1'' and ``RF2'' in Figure~3(a3) and (a5)) 
in the filament near the southern endpoint and traced their motions, which are displayed in Figure~4.
The dotted lines in Figure~4(a1)--(a3) represent the profiles of the same filament thread at different times.
RF1 is chosen at the intersection point between the dotted line and the circle, along which RF1 turned around the center.
RF1 rotated about 72\degr\ in 6 minutes, which derives an average angular speed of 12\degr\ min$^{-1}$.
As for RF2, it circled $\sim$32\degr\ around the center from 02:57:52 to 03:01:30 UT,
which outputs an average angular veloctiy of $\sim$9\degr\ min$^{-1}$.
If we take the mean value of the two average angular velocities, i.e. 10.5\degr\
min$^{-1}$, as the average untwisting motion speed of the filament, then the peak value of
the stored twist can be derived by 10.5\degr\ min$^{-1}$ times the entire rotation duration 23 min (from 02:40 to 03:03 UT).
The estimated total twist is $\sim$240$\degr$ or 1.3$\pi$, which is only about half of 
the threshold 2.5$\pi$ (the lower limit of $\Phi_{c}$) for the occurrence of kink instability. 
According to the result, it is hard to conclude that kink instability triggered the filament eruption.

\subsection{Photospheric Flux Evolution} \label{subsec:}
Essentially, solar eruption results from the loss-of-equilibrium of a magnetic system due to an imbalance between magnetic tension and compression (Forbes \& Isenberg 1991).
Sometimes, this situation can be caused by the interaction between the erupting structure
and its nearby magnetic field, for example, the aforementioned flux cancellation (Chen \& Shibata 2000; Lin et al. 2001).
Thus, we also investigated the photospheric magnetic flux evolution associated with this event (see Figure~5, 6 and the movie~3).
The HMI magnetogram in Figure~5(a) shows the photospheric longitudinal fields of the AR and its spatial relation to the filament projected in the surface.
We can see that the positive field of the AR was more concentrated than the negative field and there was no obvious magnetic neutral line in the AR. 
The two ends of the filament were separately anchored onto the northern positive flux and the negative flux in a southern remote area.

We focused on the flux changes in the region close to the filament northern end (Figure~5(b1)--(b5)), since some small-scale activities (EUV brightenings) had taken place there before the eruptions.
The HMI data clearly reveal that some negative fluxes started to be enhanced near the filament footpoint from $\sim$22:00 UT and reached its maximum at $\sim$23:15 UT on November 9.
Then, they moved westward and collided with the neighboring positive fluxes, as indicated by the arrows in Figure~5(b2).
Until 02:05 UT on November 10, i.e. just prior to Erup1, these moving fluxes were almost completely cancelled by the opposite polarity fluxes and disappeared.
We also calculated the positive and unsigned negative longitudinal magnetic fluxes in the cancellation region (marked by the box in Figure~5(b2)) from 21:00 UT on November 9 to 03:00 UT on November 10.
The time variations of the fluxes are presented in Figure~5(c).
It can be seen that the unsigned negative flux apparently mounted from $\sim$22:00 UT 
and reached the peak value of $\sim$6 $\times$ 10$^{18}$ Mx at $\sim$23:15 UT on November 9.
Since then, it dropped gradually with small-amplitude oscillations and the positive flux declined as well.
Until Erup1, the negative flux had been decreased to a less level than before it augmented on November 9.
The flux-time curves likewise reflect the flux enhancement and cancellation processes occurring near the filament end during the period of 5 hr prior to the eruption.

The flux cancellations presented above are probably caused by the slow magnetic reconnections between the moving negative fluxes and its nearby positive fluxes (Wang \& Shi 1993), which would further lead to some small-scale activities observed as the EUV brightenings on some occasions.
The high degree of temporal and spatial correlations between the flux cancellations and the filament evolution make us believe that there is an intimate connection between these activities.
It is very likely that the interactions between the converging and collisional fluxes at the end of the filament firstly affected the mechanical equilibrium of the filament, then resulted in the filament activations and disturbances and finally triggered the following explosions.

Figure~6 shows the close-up of the flux convergence and cancellation region with the HMI vector magnetograms.
It can be found that some transverse fields between the cancelling fluxes, as marked by the circles, were strongly sheared.
The strength of the strongest sheared transverse fields reaches $\sim$300 G and the shear angle is nearly 90\degr.
Although it is hard to clarify the connectivities of the associated magnetic loops, these sheared fluxes strongly suggest that the cancelling fluxes, at least part of them, do not come from the footpoints of a single loop.
Sheared fields are often observed in solar ARs (e.g., Wang et al. 2002; Schrijver et al. 2005; Zhang et al. 2007; Sun et al. 2012; Chen et al. 2015).
According to the scenario described by Moore et al. (2001), the strongly-sheared core fields are favorable to the occurrence of the tether-cutting reconnection, which would give rise to the eruption by weakening the magnetic tethers holding the filament down.
The flux cancellation model proposed by van Ballegooijen \& Martens (1989) also suggests a similar physical process.
In this event, the sheared transverse fields presented by the vector field data further indicate the important role of the flux cancellation in causing the eruptions.
Using the vector magnetograms without resolution of the 180\degr\ ambiguity, Wang \& Shi (1993) and Zhang et al. (2001) have found that the opposite polarities in the cancelling magnetic features are not the footpoints of a single flux loop, but the footpoints from two separated loops respectively, and strong shear would develop at the interface of the opposite polarities.
Our observations confirm their results and likewise imply the reconnection essence of the flux cancellation.



\section{Summary and Discussion} \label{sec:summary}
We study the kinematics and morphology of an erupting filament during its two successive confined eruptions (Erup1 and Erup2) in a decaying AR.
Our observations unambiguously present evidence for the confinement of the eruption from the overlying magnetic arcade and the rotational motion of the erupting filament due to unwinding, suggestive of a preceding twisted configuration of the filament.
The total stored twist of $\sim$1.3$\pi$ is obtained by analyzing and calculating the rotational angles of the filament during its untwisting, which is far from satisfying the criteria for kink instability.
We note that the photospheric longitudinal fluxes around the filament underwent conspicuous changes, especially the flux convergence and cancellation in the vicinity of the northern end of the filament, during the period of several hours prior to the eruptions.
Strongly-sheared transverse fields are also found in the cancelling magnetic features from the vector magnetograms.
These results support the flux cancellation model (e.g., van Ballegooijen \& Martens 1989), in which the interaction between the converging and sheared fluxes of opposite polarity gradually destroys the balance between the magnetic tension and compression of the filament system, activate the filament and trigger the final ejection.
It is worth noting that we are not claiming that the erupting-filament behavior that we observe in the present event is typical of all eruptions, and we expect that the evolution of filaments during various eruptions can show a variety of patterns.
Here, we would like make a comparison with the observations of Yan et al. (2014a) and (2014b).
In their cases, the erupting filaments also exhibited untwisting motions, suggesting the structures of flux rope. 
Different from ours, their results showed that the total twists of the filaments were separately $\sim$3$\pi$ and 5$\pi$, exceeding the typical threshold (2.75$\pi$; T\"{o}r\"{o}k \& Kliem 2003) for kink instability.
Consequently, they proposed the kink instability as the triggering mechanism of their studied eruptions. 

It should be pointed out that we only observed the untwisting motion of the filament during Erup2.
Although the NVST observations with the better spatial resolution did not cover Erup1, any rotational motion or kinked structure was not detected during the first eruption from the AIA intensity images.
Hence the twist stored in the filament and released during Erup2 was perhaps formed during Eurp1, which is in good agreement with the appearance of the coronal sigmoidal structure in the meantime (Figure~2).
The main process of this event under our study may happen like this scenario: at first, under the influence of the flux cancellation, the filament was gradually destabilized accompanied by the EUV brightenings and was finally ejected to form Erup1 when the system reached a point that no nearby equilibrium is accessible; during Erup1, perhaps due to the reconnection of sheared fields or the injection of the poloidal field, some twist was stored into the filament-carrying field, which is corresponding to the production of the coronal sigmoid; then the twist was released in the following eruption observed as the untwisting of the filament during Erup2.
Allowing for a ``double-decker'' configuration of filament or flux rope (e.g., Liu et al. 2012; Li \& Zhang 2013; Cheng et al. 2014a; Kliem et al. 2014; Zhu \& Alexander 2014; Dhakal et al. 2018; Tian et al. 2018) or even a mutil-flux-rope system (e.g., Awasthi et al. 2018; Hou et al. 2018), there is a second possibility besides the situation described above. It might be that there are two sets of erupting fields, one on top of the other. The upper field erupts to cause the first eruption, and the lower field erupts to cause the second eruption. The lower field could contain twist, but it is being held down by a less-twisted upper field. The flux cancellation results in the overlying top field erupting outward (Erup1), which is not strong enough to escape as a CME. During this eruption, the filament and the field containing it rose up, and also moved toward the south roughly following the AR's PIL, that caused new brightenings to occur along the magnetic neutral line. Since the neutral line naturally has an approximate sigmoidal shape, the additional brightenings made the overall structure appear as a sigmoid in 211 \AA. Then, with the upper field removed, the lower, more twisted, field had a chance to erupt upward and unwind (Erup2). Again it did not have enough energy to escape as a CME, and so it too remained as a confined eruption.

\acknowledgments
We thank the referee for constructive comments which were helpful in improving the paper. The data are used courtesy of NVST and $SDO$ science teams. 
This work is supported by NSFC (11790304, 11533008, 11790300, 11673034, 11673035, 11773039), and Key Programs of the Chinese Academy of Sciences (QYZDJ- SSW-SLH050).

\clearpage
\begin{figure}
\epsscale{1}
\plotone{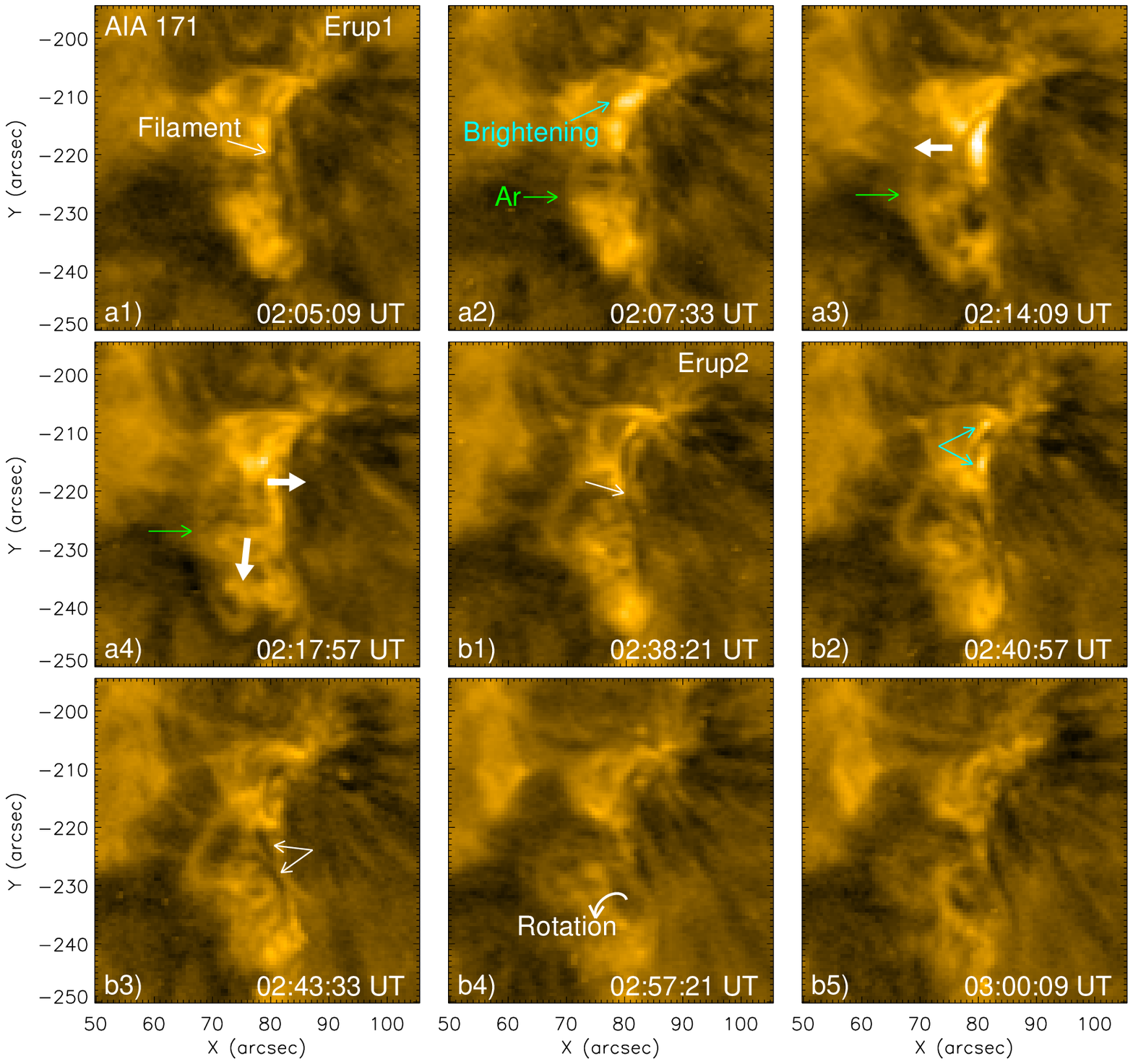}
\caption{AIA 171 \AA\ images showing the evolutions of the filament during Erup1 ((a1)--(a4))
and Erup2 ((b1)--(b5)) (also see the movie~1).
The ``Ar'' in panel (a2) means the overlying arcade above the filament and the green arrows in panels (a2)--(a4) point out the locations of Ar in the course of Erup1.
The thick arrows in panels (a3) and (a4) denote the movement directions of the filament 
during Erup1. 
The white and turquoise arrows in panels (b1) and (b2) point to the filament and the EUV brightenings prior to Erup2, respectively.
The arrows in panel (b3) indicate the bifurcation of the filament during Erup2.
\label{fig1}}
\end{figure}

\begin{figure}
\epsscale{1}
\plotone{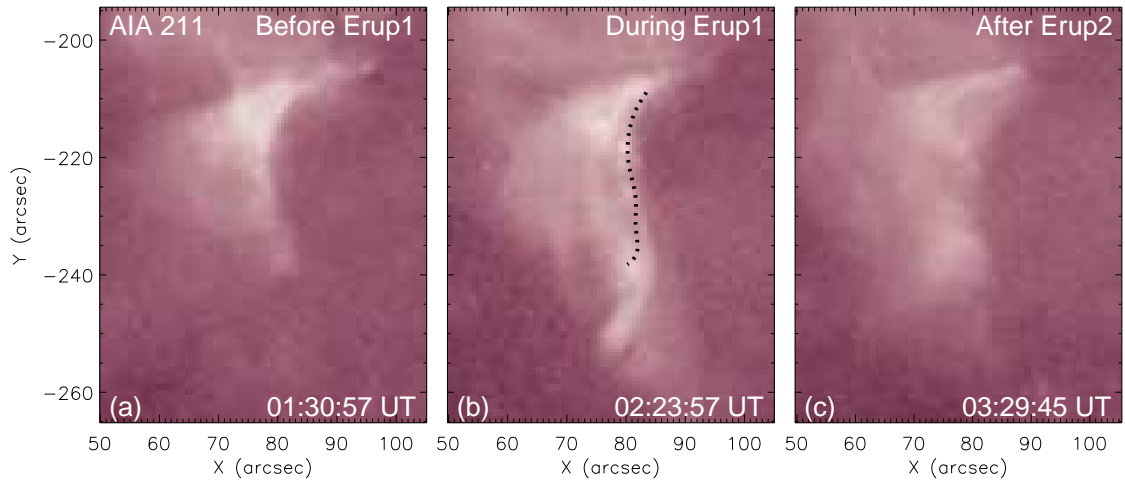}
\caption{AIA 211 \AA\ images displaying the changes of the coronal structure in the AR before (a), during (b) Erup1 and after Erup2 (c) (also see the movie~1).
The dotted line in panel (b) indicates the location of the filament before eruption.
\label{fig2}}
\end{figure}

\begin{figure}
\epsscale{1}
\plotone{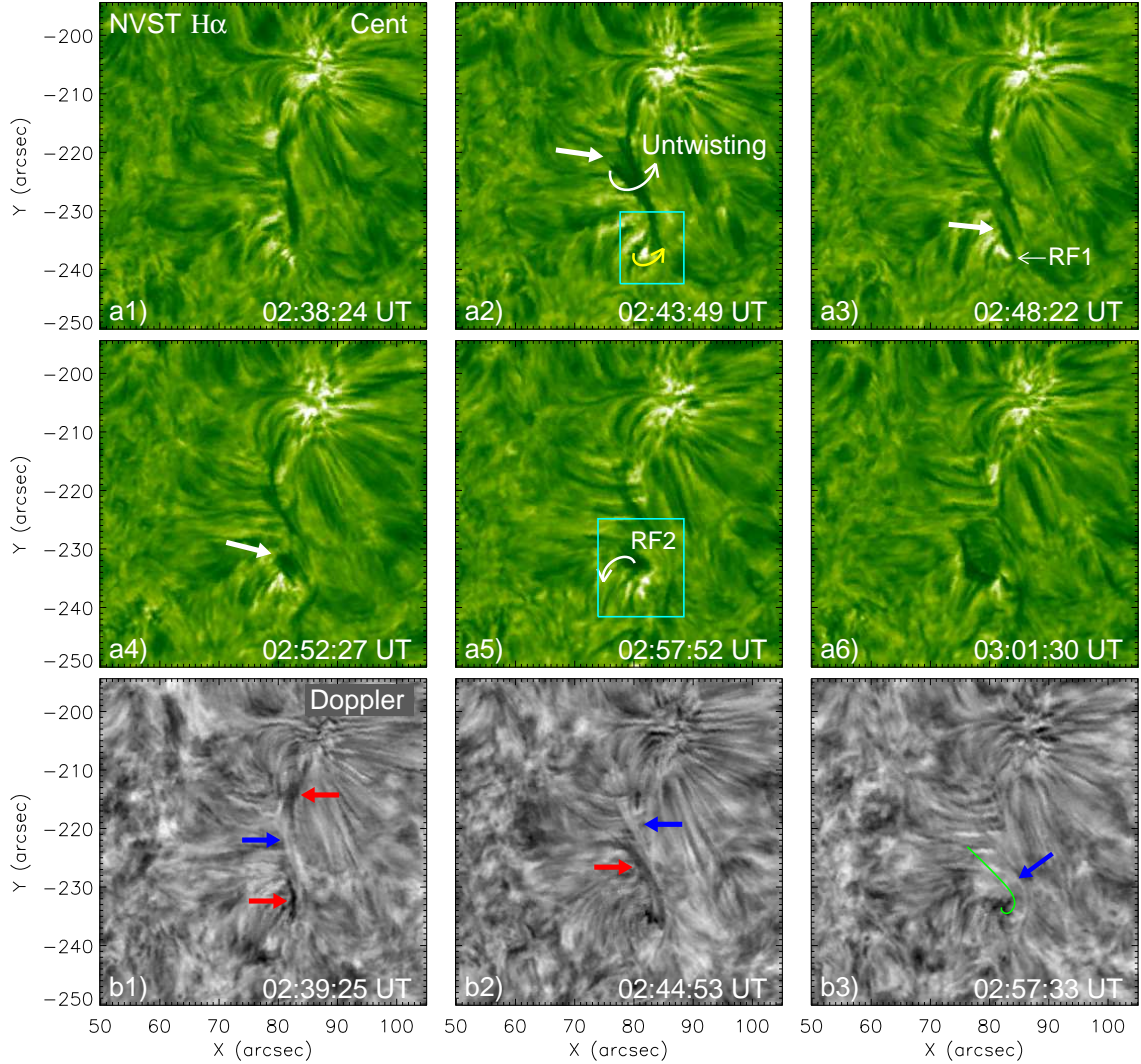}
\caption{NVST H$\alpha$ line center ((a1)--(a6)) and Doppler ((b1)--(b3)) 
intensity data showing the evolution of the filament during Erup2 (also see the movie~2). 
The thick arrows in panels (a2)--(a4) aim at the bifurcated structure of the filament due to 
untwisting during Erup2.
The curved arrows in panels (a2) and (a5) denote the rotation of the filament 
at its southern root.
The rectangles in panels (a2) and (a5) indicate the field of view (FOV) of the top and bottom panels of Figure 4, respectively.
The red and blue arrows in panels (b1)--(b3) point to the red- and blueshifted parts 
of the filament, respectively.
The curve in panel (b3) indicates the helical structure of the filament rooted in 
its southern endpoint.
\label{fig3}}
\end{figure}

\begin{figure}
\epsscale{1}
\plotone{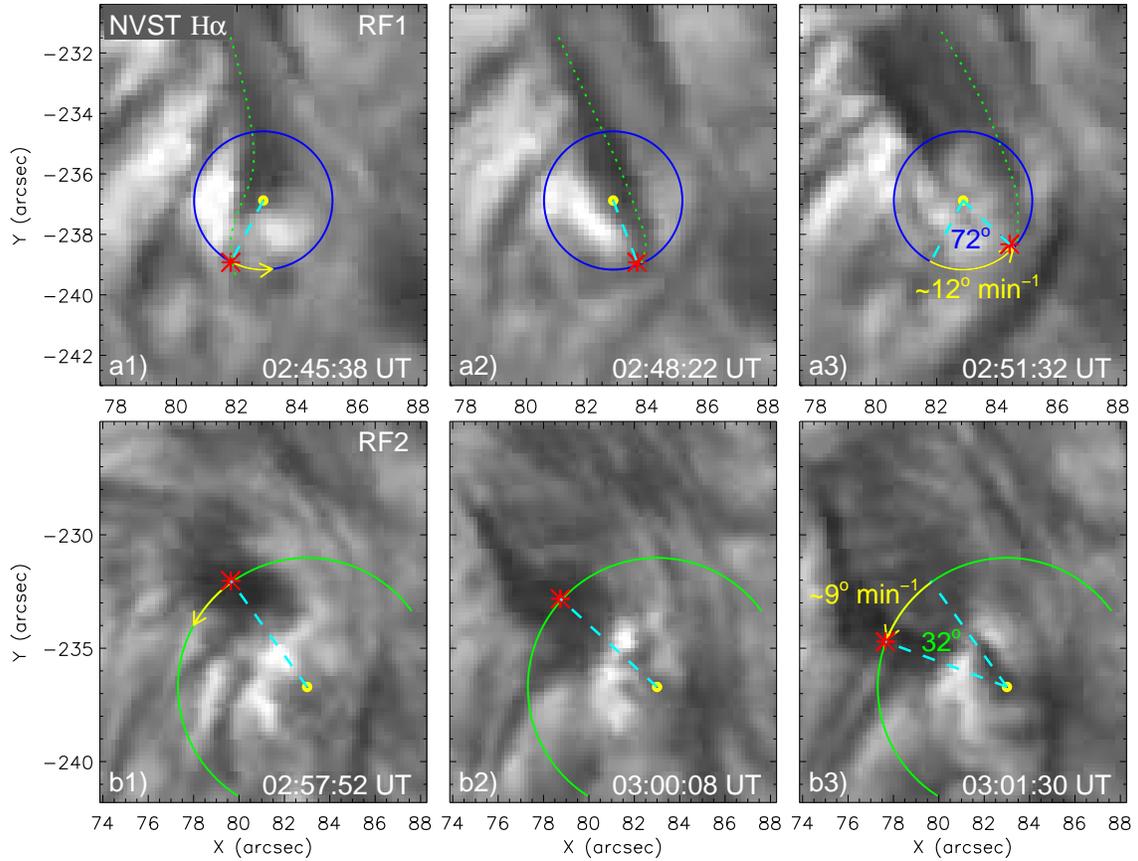}
\caption{Two rotating features RF1 ((a1)--(a3)) and RF2 ((b1)--(b3)) in the filament presented by NVST H$\alpha$ data.
The dotted lines in panels (a1)--(a3) indicate the same thread of the filament.
The asterisks in panels (a1)--(a3) and (b1)--(b3) correspond to the positions of 
RF1 and RF2, respectively.
The yellow arrows in panels (a3) and (b3) separately denote the trajectories of 
RF1 and RF2.
\label{fig4}}
\end{figure}

\begin{figure}
\epsscale{1}
\plotone{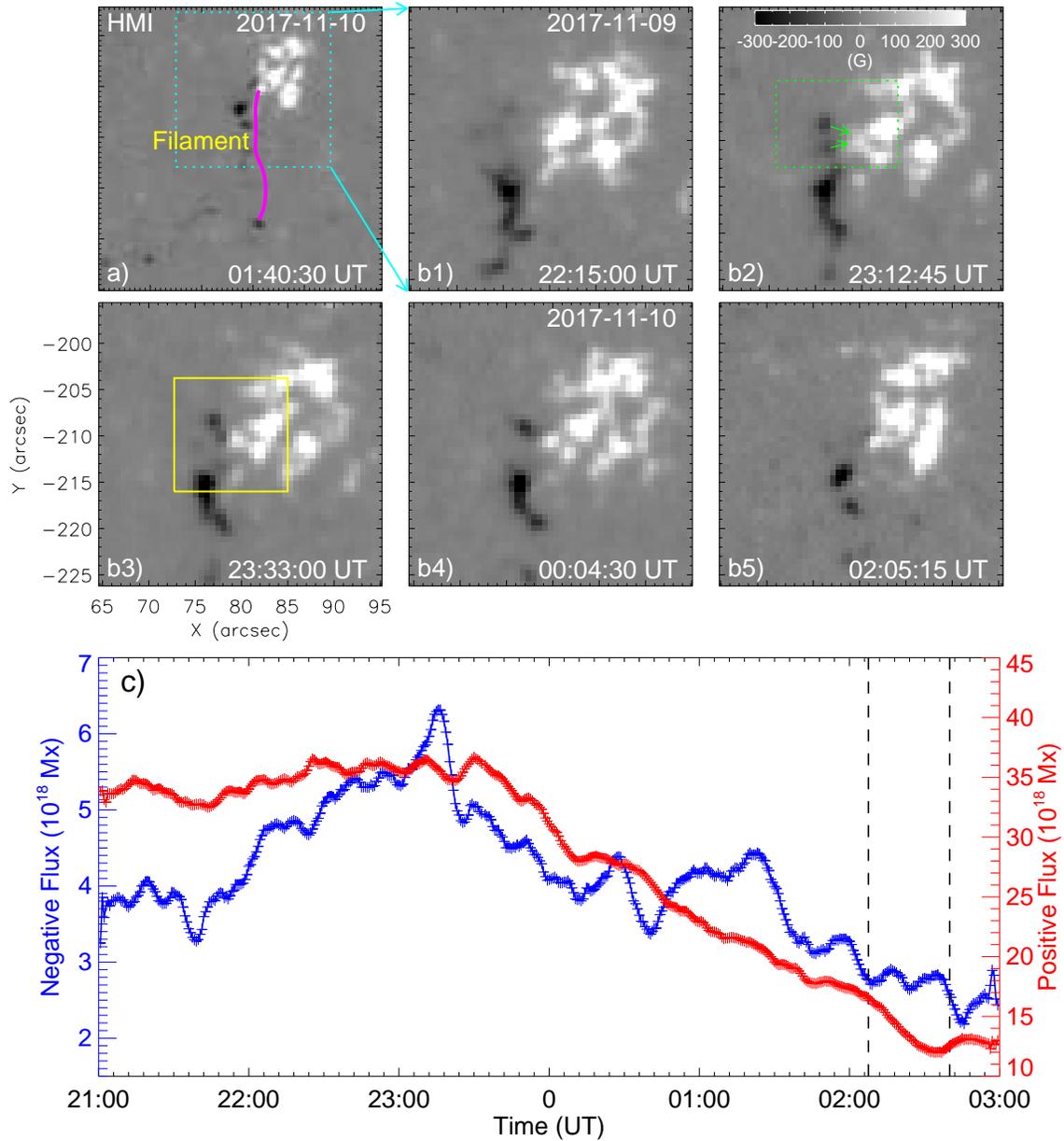}
\caption{((a)--(b5)) HMI longitudinal magnetograms (also see the movie~3); 
(c) time profiles of the positive (red plus) and unsigned negative (blue plus) 
longitudinal fluxes in the area indicated by the box in panel (b2).
The boxes in panels (a) and (b3) separately represent the FOVs of 
panels (b1)--(b5) and Figure~6.
The arrows in panel (b2) suggest the motion directions of the newly emerged 
negative fluxes near the northern end of the filament.
The dashed lines in panel (c) correspond to the onset times of Erup1 and Erup2.
\label{fig5}}
\end{figure}

\begin{figure}
\epsscale{1}
\plotone{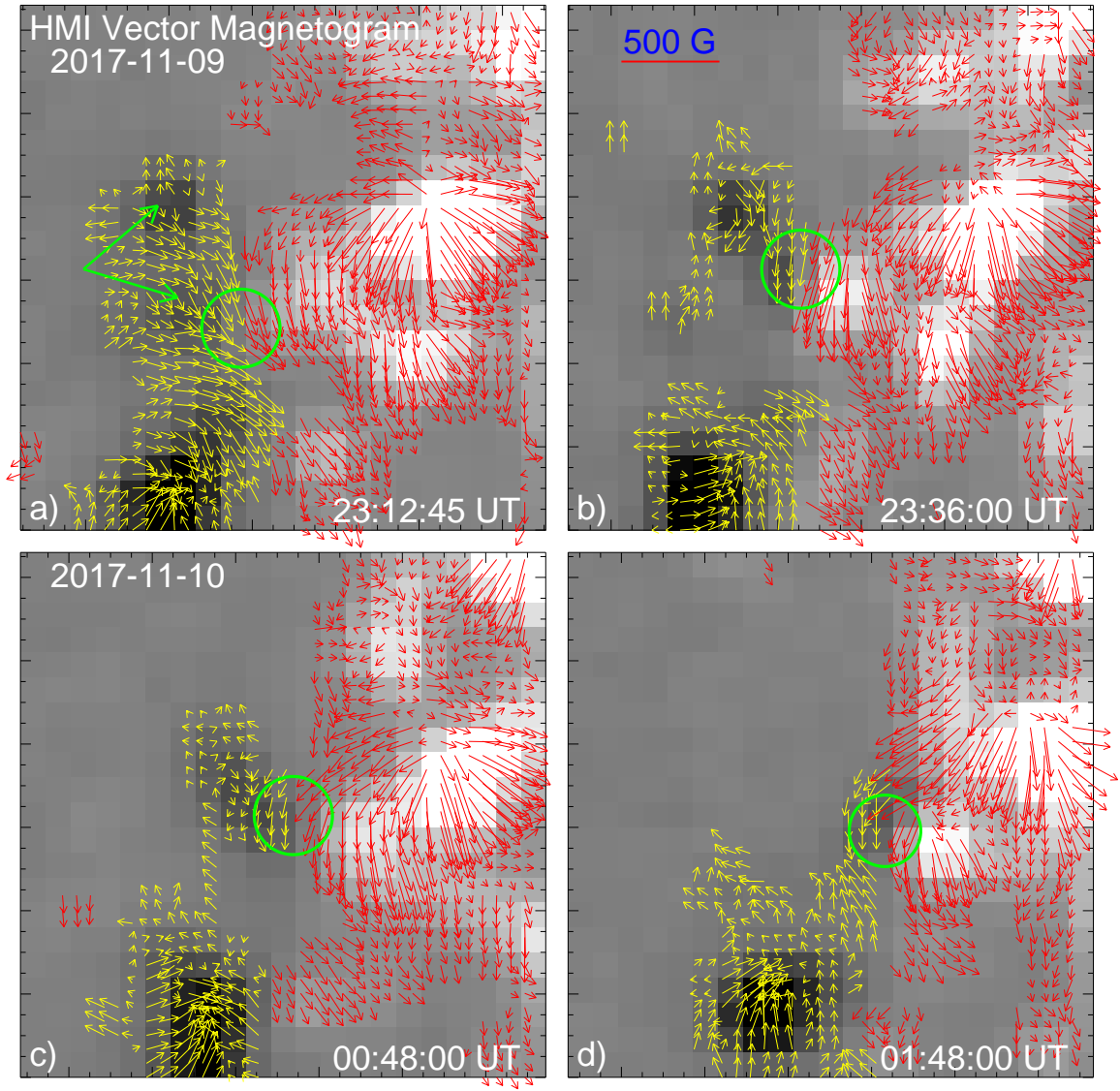}
\caption{HMI vector magnetograms;
The FOV is denoted by the rectangle in Figure~5(b3).
The green arrows in panel (a) point to the newly emerged negative fluxes.
The circles mark the areas between the cancelling fluxes where sheared magnetic fields appeared.
\label{fig6}}
\end{figure}


\end{document}